\documentstyle[prb,aps]{revtex}

\tighten
\begin{document}
\draft

\title{
Nonperturbative approach to the Hubbard model in $C_{60}$ cluster}

\author{J. Gonz\'alez}
\address{Instituto de Estructura de la Materia, Serrano 123,
28006 Madrid, Spain}
\author{J. V. Alvarez}
\address{Escuela Polit\'ecnica Superior, Universidad Carlos III,
Butarque 15, Legan\'es, 28913 Madrid, Spain}

\date{\today}
\maketitle

\begin{abstract}

We propose a computational scheme for the Hubbard model in the
$C_{60}$ cluster in which the interaction with the Fermi sea of
charges added to the neutral molecule is switched on
sequentially. This is applied to the calculation of the balance
of charging energies, within a low-energy truncation of the
space of states which produces moderate errors for an
intermediate range of the interaction strength.

\end{abstract}

\pacs{36.40, 71.10}

   The description of the electronic properties of alkali-doped
fullerene crystals poses a challenge to the traditional methods
of condensed matter physics, as the properties of these systems
depend on the conjunction of the large scale structure of the
crystal with the small scale effects inside each fullerene. This
conjunction is probably responsible for some of the exotic
phenomena observed in these materials, the most important of
which is the relatively high-$T_c$ superconductivity of the $A_3
C_{60}$ compounds, $A$ being an alkaline metal\cite{htc}.
It is worth to
remind that the transition temperatures attained in some of the
compounds are above the 40 K, while other carbon based
materials like the intercalated compounds of graphite are
superconducting up to $\sim$ 1 K. Most part of the theoretical
frameworks proposed for the explanation of the high-$T_c$
superconductivity rely on specific properties of the
electron-phonon interaction\cite{ep,vib,pietr,jt,bj}
or the electron interaction\cite{chk,flr,lr} inside the $C_{60}$
cluster.

  In this paper we undertake a formal study of the electronic
interaction in the $C_{60}$ molecule, which is perhaps the worst
understood from the phenomenological as well as from the
theoretical point of view. In the first place, it seems that,
even taking into account screening effects due to the
polarization of neighboring molecules, the Coulomb
interaction should give rise to a significant repulsion ($\sim$
1 eV)
between added charges in the $C_{60}$ molecule\cite{ant,coul}.
There have been some
attempts to explain that, in spite of the bare strong repulsive
interaction, the effective interaction between charges over the
whole cluster could have a completely different character,
because of strong dressing effects\cite{chk}
or the existence of selection
rules for the transitions between different charged states\cite{nos}.
 From the theoretical point of view, the properties of the $C_{60}$
cluster are difficult to describe since the molecule has an
intermediate scale which renders rather useless the standard
many-body methods while a numerical diagonalization approach is still
unfeasible. In the present paper we deal with a Hubbard model
for the electron interaction and propose an
alternative to perturbation theory for the same, which may
also give the correct physical picture in the strong coupling
regime.

  The nonperturbative approach we propose is based on the idea of
dealing with a reduced Fermi sea comprised of states down the
highest unoccupied molecular orbital, up to a given level. One
introduces in this process a truncation of the whole set of
one-particle occupied orbitals and the way to get under control
this approximation has two steps. The first consists of
enlarging the set of states in the Fermi sea by adding orbitals
of lower energy, up to a point in which the observables
considered get stabilized. In practice this happens when the
many-body system has reached a too large number of states to
admit an exact diagonalization of the hamiltonian. Then one has to
introduce an approximation method which still renders the
problem numerically solvable while producing accurate enough
results. We implement in this paper an iteration in the number
of allowed particle-hole excitations from the reduced Fermi sea.
Thus we may be dealing with a Fermi sea made of 28 states, for
instance, and considering an iteration $n = 0, 1, 2, \ldots$ of
respective many-body spaces of states with up to $n$
particle-hole excitations from the 28 occupied orbitals. The
first terms of the succession still lead to a truncated space of
states in which the interaction can be numerically solved, but
the important point is that the convergence of quantities like
the ground state of the system is exponential with regard to
$n$. In any event, the error introduced by stopping the
iteration at a given number $n$ can be estimated. Comparing this
computational method with perturbation theory, it turns out that
diagonalizing the system with up to $n$ particle-hole
excitations already accounts for a partial sum of the usual
perturbative series, since it amounts to consider all the
diagrams which can be cut in two parts showing up to $2n$
internal electron lines.

  The crucial point in our computational scheme lies on the
progressive enlargement of the Fermi sea and the possibility of
keeping under control the variations of the observables
computed. In this sense, our nonperturbative approach is
particularly well-suited in the calculation of quantities which
do not depend extensively on the size of the Fermi sea, as is
the case of the balance of charging energies or correlation
functions in the molecule. In the present situation we focus on
the evaluation of the energy needed to place a pair of electrons
over a neutral $C_{60}$ cluster, measured with respect to a
configuration of a single electron in each of two different
molecules. This is what has been called $E_{pair}$ in ref.
\onlinecite{chk}, and it has been linked to the signature of a purely
electronic mechanism for the superconductivity in the
alkali-doped materials. For the sake of simplicity we will
consider a Hubbard model of interaction for the 60 $\pi$
orbitals of the $C_{60}$ cluster

\begin{equation}
H = -t \sum_{i,j,\sigma} a^{+}_{i \sigma} a_{j \sigma} +
 U \sum_{i} n_{i \uparrow} n_{i \downarrow} \;\;\;\;\;\;\;\;\;\;
\sigma = \uparrow , \downarrow
\label{ham}
\end{equation}

where the first sum is over nearest neighbors $i,j$ and $n_i$ is
the electron number operator at site $i$. The spectrum of
one-particle energy levels in the free theory has the well-known
pattern shown in Fig. \ref{one} \cite{hbr}.
In the neutral $C_{60}$ molecule, all
the levels are filled up to the quintuplet right below zero
energy. The charged states of the molecule correspond to filling
the positive energy levels, starting from the first triplet
above zero energy. The states in the triplet transform under
spatial rotations as those of an $l = 1$ angular momentum
representation\cite{mp}.
Thus, the  $C_{60}$ molecule with one added charge
must have a six-fold degenerate ground state, corresponding to
spin $s = 1/2$ and angular momentum $l = 1$ quantum numbers.
This degeneracy remains unchanged in the interacting theory,
since both quantities are conserved in the Hubbard model
(\ref{ham}). In the same fashion, the $C^{--}_{60}$ anion would
have in the free theory a 15-fold degenerate ground state,
comprising a multiplet with total spin $s = 1$ and total angular
momentum $l = 1$, another with $s = 0$, $l = 2$, and a singlet
state with $s = 0$, $l = 0$. The degeneracy between these three
multiplets is removed, though, in the interacting theory.

  The parameters which may apply to the actual description of the
$C_{60}$ cluster are $t \sim 2$ eV and $U \sim 5 - 10$ eV. It
seems adequate to consider the interaction between
the states in the triplet as the starting point of our
computation, since the energy difference with respect to the
lower one-particle multiplets is more than 1.5 eV. Thus,
neglecting the interaction with the rest of occupied levels
below the triplet, the multiplet with $l = 1, s = 1$ would have
the lowest energy and $E_{pair} = 0$, as it is possible to place
the two electrons with parallel spins without feeling the
on-site repulsion. For the above range of parameters it is easy
to check that the hybridization with higher unoccupied levels is
insignificant, except for the states of the second triplet above
zero energy. Therefore, level crossing involving the ground
state of the doubly charged molecule and a negative value of
$E_{pair}$ can only be due to the mixing between the two
triplets and the interaction with the negative energy levels.
We will consider these as our Fermi sea,  whose dynamics we want
to include progressively.

  As a first step we consider a many-body space of states in which
the two triplets above zero energy hybridize and only
particle-hole excitations from the quintuplet right below zero
energy are included. In this case, the dimension of the space of
states for the different anions is small enough that the respective
ground states can be obtained by numerical diagonalization. The
results are summarized in Table \ref{t1}
and plotted in Fig. \ref{two}, for
$t = 1.8$ eV and an on-site repulsion ranging from 2.5 eV to
10.0 eV. For the $C^{--}_{60}$ anion we give the energy
corresponding to the singlet $l = 0, s = 0$, as well as that of
the level with quantum numbers $l = 1, s = 1$. In all the
instances it can be seen that the latter has the lowest energy,
keeping $E_{pair} = E^{(2)} + E^{(0)} - 2E^{(1)}$ well below
0.01 eV ---the maximum value we find is for $U = 10.0$ eV,
$E_{pair} \approx 0.001$ eV. The results for $E_{pair}$ match at
small $U$ with those from perturbation theory, which predicts
for the singlet a slope equal to 1/20 at the origin\cite{chk}.
Our fit of
the points in the figure gives in turn $\approx 0.0499$. It is also
instructive, for later use, to check the convergence to the
correct ground state energies as the space of states is
constrained each time to contain states with a number of
particle-hole excitations $\leq n = 1, 2, 3, \ldots $, up to the
total number of particles in the Fermi sea. The screening due
to these polarization effects is enhanced at even values of $n$,
but the excitation of two additional particles produces a much
weaker effect as $n$ increases. If we call $E(n)$ the
approximation to the ground state energy $E_{\infty}$ when
taking into account states with up to $n$ particle-hole
excitations, we find that the $n$-dependence may be reasonably
fitted by

\begin{equation}
E(n) = E_{\infty} + b e^{-a n}
\label{for}
\end{equation}

with $a \sim O(1)$. We conjecture that the value of $a$ does not
depend significantly on the size of the Fermi sea, while it
certainly is a decreasing function of $U/t$.

  Next we proceed to enlarge the Fermi sea by allowing also
particle-hole excitations from the multiplet below the
quintuplet considered before. The many-particle space of states
we are dealing with now is built out of the  two hybridized
triplets and particle-hole excitations from the 5-fold and
9-fold degenerated multiplets below zero energy. This space has
such a large dimension that makes unfeasible the numerical
diagonalization of the hamiltonian. We may estimate, however,
the screening effects of the enlarged Fermi sea by constraining
the maximum number $n$ of allowed particle-hole excitations.
Again these effects are enhanced for even values of $n$, and the
truncation at $n = 2$ still gives rise to a manageable space of
states. The ground state energies computed in this approximation
for the neutral, singly and doubly charged molecules are given
in Table \ref{t2}. The respective differences $E(n = 2) - E(n = 0)$ can
be evaluated, showing that the  value of $b$ in (\ref{for}) is
now between 4 and 4.5 times greater than for the smaller Fermi
sea considered before. The empirical law (\ref{for}) enables
then to estimate the error produced by stopping at $n = 2$. This
is anyhow a systematic error which goes in the same direction
for $E^{(0)}, E^{(1)}$ and $E^{(2)}$, propagating with a weaker
influence to a quantity like $E_{pair}$. The approximated values
of $E_{pair}$ from the above truncation together with the error
estimates are plotted in Fig. \ref{two}. The fit of the points in the
figure gives a slope at the origin $\approx 0.053$, in good
agreement with perturbation theory.

  The inspection of the respective curves of $E_{pair}$ in Fig.
\ref{two} for the Fermi sea with 10 states and for that with 28 states
(including spin) shows a clear stabilization of the results in the
second case. One should not expect a significant modification of
the curves by the inclusion of lower energy multiplets in the
Fermi sea since, moreover, particle-hole excitations from the
low one-particle levels would require higher energy. It is
therefore plausible that the physical picture which emerges from
Fig. \ref{two} is essentially correct. The screening effects which
take place within the $C_{60}$ molecule turn out to be very
efficient reducing the bare electronic repulsion between added
charges in the $s = 1$ multiplet. This repulsion is appreciably
smaller, already at intermediate values of $U/t$, than that
estimated by perturbation theory. The doubly charged state with
$s = 0$ seems to be always an excited state, opposite to what
happens in smaller clusters like the truncated
tetrahedron\cite{sr}. The
reason for this different behaviour is that, switching off the
interaction with the Fermi sea in the $C_{60}$ cluster, the
interaction in the partially occupied triplet already places the
singlet state at a fair energy above the $s = 1$ multiplet and
the screening effects are not able to modify this trend. In
the cluster of the truncated tetrahedron, the interaction within
the triplet keeps the $s = 1$ level degenerated with a $s = 0$
doublet, which is driven at a lower energy by the particle-hole
excitations. It would be interesting to understand the factors
which may alter the relative position of the levels. In this
sense, a most interesting proposal has been made in ref.
\onlinecite{chklr} considering the frequency dependence of the
screening effects, which may favor energetically the $s = 0$
singlet state. The
nonperturbative framework presented in this paper may be useful
in the study of these problems, as well as in the study of other
effects like charge and spin correlations.

We want to thank F. Guinea for encouragement and useful discussions
during the development of this work.

\begin{figure}
\caption{One-particle spectrum of the $C_{60}$ cluster. Energy
eigenvalues are plotted in the horizontal axis and the multiplet
degeneracy is given along the vertical direction.}
\label{one}
\end{figure}

\begin{figure}
\caption{Plot of the values of $E_{pair} = E^{(0)} + E^{(2)} - 2
E^{(1)}$ corresponding to the data in Table
\protect\ref{t1} (points over the
solid lines) and to those in Table
\protect\ref{t2} (points over the dashed
lines). The latter are affected by the error bars mentioned in
the text.}
\label{two}
\end{figure}

\newpage
\begin{table}[h]
\centering
\begin{tabular}{|d|d|d|d|d|}
    &             &             &            &            \\
$U$ (eV) & $E^{(0)}$ (eV)  & $E^{(1)}$ (eV)  &
           $E^{(2)}$ (eV) & $E^{(2)}$  (eV) \\  \hline
    &              &             &  s = 1     & s = 0  \\ \hline
2.5 & 1.021        & 2.593       & 4.166      & 4.266   \\
5.0 & 2.007        & 3.795       & 5.583      & 5.744 \\
7.5 & 2.963        & 4.970       & 6.977      & 7.170 \\
10.0& 3.894        & 6.121       & 8.349      & 8.558 \\
\end{tabular}

\caption{Respective ground state energies of the neutral, singly
and doubly charged $C_{60}$ molecule, computed with a Fermi sea
of 10 states. The origin of energies is taken at the highest
occupied one-particle level of the neutral molecule.}
\label{t1}
\end{table}

\newpage
\begin{table}[h]
\centering
\begin{tabular}{|d|d|d|d|d|}
    &             &             &            &            \\
$U$ (eV) & $E^{(0)}$ (eV)  & $E^{(1)}$ (eV)  &
           $E^{(2)}$ (eV) & $E^{(2)}$  (eV) \\  \hline
    &              &             &  s = 1     & s = 0  \\ \hline
2.5 & $-$4.308     & $-$2.352    & $-$0.394   &  $-$0.300  \\
5.0 & 3.621        & 6.191       & 8.765      & 8.909 \\
7.5 & 11.451       & 14.643      & 17.846     & 18.018 \\
10.0& 19.218       & 23.036      & 26.872     & 27.060 \\
\end{tabular}

\caption{Respective ground state energies of the neutral, singly
and doubly charged $C_{60}$ molecule, computed with a Fermi sea
of 28 states and the approximation mentioned in the text. The
origin of energies is as in the preceding table.}
\label{t2}
\end{table}

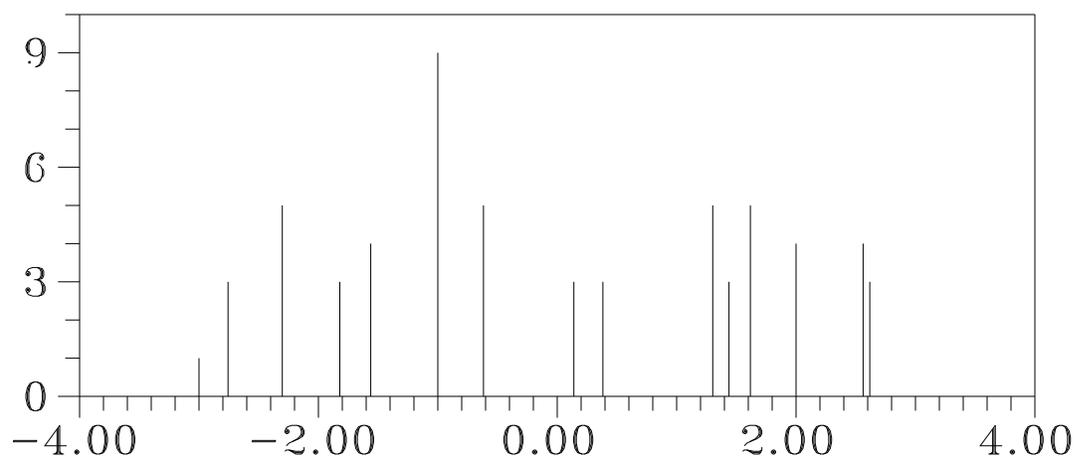
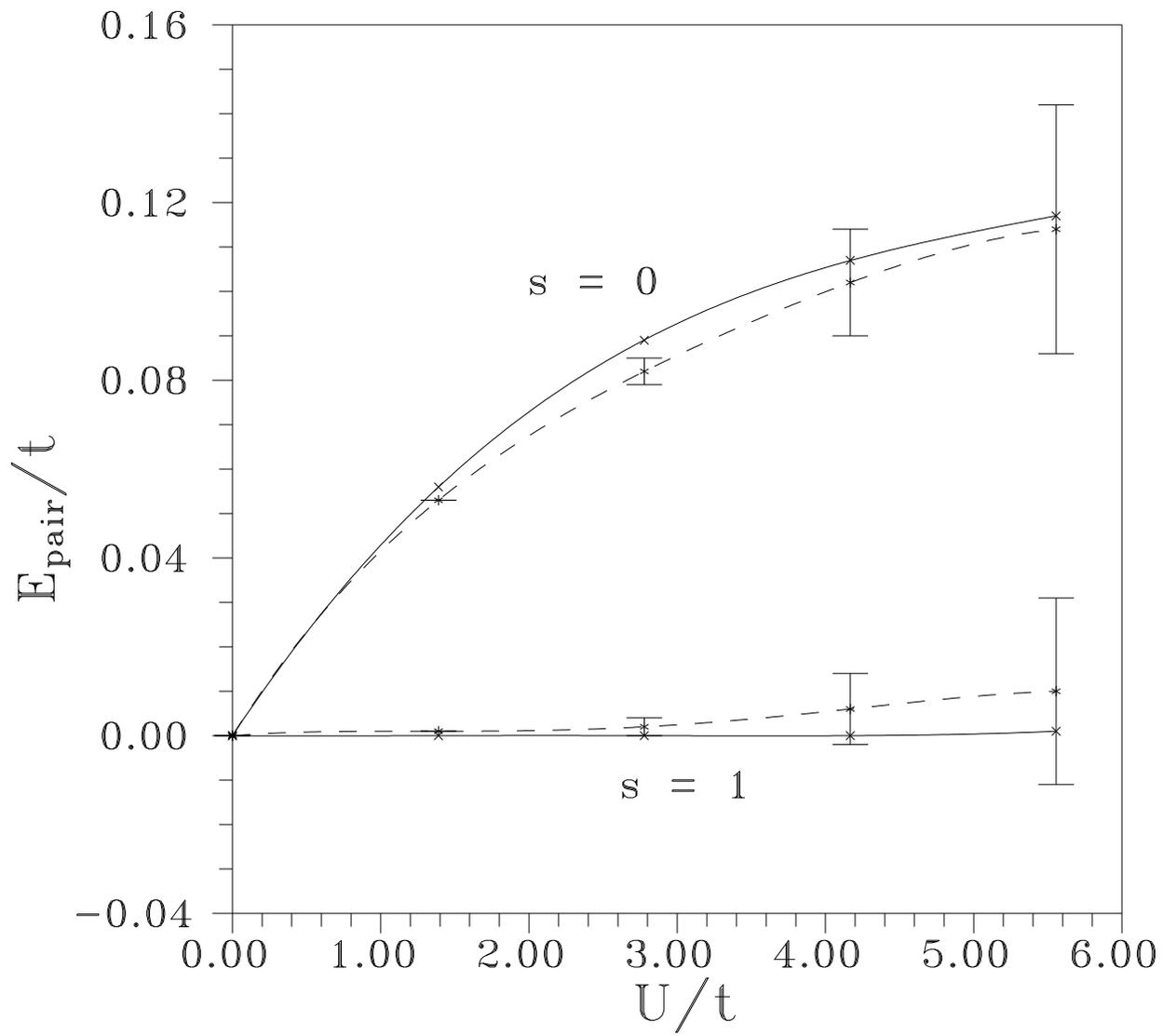
\end{document}